\documentstyle[12pt,times]{article}
\newlength{\abstractwidth}
\font\zfont = cmss10 
\newcommand{\starttext}{
\setcounter{footnote}{0}
\renewcommand{\thefootnote}{\arabic{footnote}}}

\newcommand\ZZ{\hbox{\zfont Z\kern-.4emZ}}
\newcommand\nup{{\nu^\prime}}
\newcommand\mup{{\mu^\prime}}
\newcommand\al{\alpha}
\newcommand\alp{{\alpha^\prime}}
\newcommand\de{\partial}
\newcommand\nn{\nonumber}
\newcommand\beq{\begin{equation}}
\newcommand\eeq{\end{equation}}
\newcommand\pr{\prime}
\newcommand\mo{{\mu_1}}
\newcommand\mt{{\mu_2}}
\newcommand\msp{{\mu_p}}
\newcommand\mpp{{{\mu_p}^\prime}}
\newcommand\mop{{{\mu_1}^\prime}}
\newcommand\mtp{{{\mu_2}^\prime}}
\newcommand{\dmup}{\partial_{\mu^\prime}}
\newcommand{\dnup}{\partial_{\nu^\prime}}
\newcommand{\dmu}{\partial_\mu}
\newcommand{\dnu}{\partial_\nu}
\def\inbar{\vrule height1.5ex width.4pt depth0pt}
\def\IC{\relax\hbox{\kern.25em$\inbar\kern-.3em{\rm C}$}}

\begin{document}

\begin{center}

\hfill  MIT-CTP-2928

\hfill hep-th/9911182

\bigskip

\bigskip
\vspace{3\baselineskip}

{\Large \bf Propagators for massive symmetric tensor and $p$-forms in 
\boldmath $AdS_{d+1}$}

\bigskip
\bigskip

{\bf Asad Naqvi}\\

\bigskip
{ \small \it Center for Theoretical Physics,

Massachusetts Institute of Technology, Cambridge, MA 02139, USA }

\bigskip

{\tt naqvi@ctp.mit.edu}

\bigskip

\vspace*{1cm}
{\bf Abstract}\\

\end{center}

\noindent
We construct propagators in Euclidean $AdS_{d+1}$ space-time for 
massive $p$-forms and massive symmetric tensors.

\bigskip

\starttext
\baselineskip=18pt
\section{Introduction} 
\setcounter{equation}{0}
Calculations of correlation functions in Type IIB supergravity on $AdS_5\times S^5$ have been performed extensively. These calculations 
study the strong coupling dynamics of ${\cal N}=4$ $SU(N)$ Yang-Mills
theory for large $N$ as a result of the AdS/CFT correspondence
conjectured in \cite{maldacena,polyakov,witten}. For calculations of 4-point
functions, bulk to bulk Green's functions, describing propagation between
two points in the interior of $AdS$, are required. Propagators for scalar
fields have been discussed in \cite{scalar},
and propagators for massless and massive gauge bosons were obtained 
in \cite{allenjacob}.
In \cite{dfmmr}, a new method for calculation of these propagators was 
discussed
in which ansatze for the bi-tensor propagators were used which 
naturally separated
the gauge invariant parts from the gauge artifacts. The gauge artifacts did
not contribute if sources of the fields were conserved currents. Thus gauge
fixing was unnecessary. 

In this paper, we use the same method as in \cite{dfmmr} but now for massive
$p$-forms and the massive symmetric tensor fields. Since the fields are 
massive, there
is no gauge invariance which guarantees that the sources are conserved. However, 
as we will see, using a similar ansatze for the $p$-forms 
(writing the propagator
as a physical part and a pure gauge) in the massive case considerably 
simplifies the calculations. 
The pure gauge part is annihilated by the Maxwell operator and just
appears multiplying $m^2$ in the equation of motion. For the massive
symmetric tensor, the pure gauge part of the ansatz corresponds to
diffeomorphisms at $z$ (where the propagator describes
propagation from $z$ to $w$). However, for our propagator to have
symmetry under the exchange $z \leftrightarrow w$, we need to
add a term which corresponds to diffeomorphisms with at $w$. This
term will not be annihilated by the ``wave operator''. However, writing
the propagator in this form still simplifies the calculation considerably. 

The paper is organized as follows. In section 2 we discuss the massless
2-form case as a warm-up exercise (this case was discussed in \cite{iosif}). 
We also find the propagators for
the massive anti-symmetric tensor. We generalize the calculation
of section 2 to $p$-forms in section 3. In section 4, we perform 
the calculation for 
the propagator of the massive
symmetric tensor. In section 5, we check the short-distance limit of 
our results and find that they match with the short distance limit of 
the corresponding propagator
in flat space. We end with a summary of our results in section 6. 

We will work in Euclidean $AdS_{d+1}$, which can be regarded as an
upper half space $z_0 > 0$ in a space with coordinates $z_0,z_1 \dots z_d$,
and metric, 
\beq
ds^2=\frac{1}{z_0^2}(dz_0^2+\sum_{i=1}^{d}dz_i^2).
\eeq
The $AdS_{d+1}$ scale has been set to unity and the metric describes a space
with a constant negative curvature $R=-d(d+1)$. We will introduce a chordal
distance $u$ in terms of which invariant functions and tensors on $AdS_{d+1}$ can
be expressed most simply: 
\beq
u \equiv \frac{(z-w)^2}{2z_0 w_0},
\eeq
where $(z-w)^2=\delta_{\mu \nu}(z-w)_\mu (z-w)_\nu$  
is the ``flat Euclidean distance". We will construct basic
bi-tensors by taking derivatives with respect $z$ or $w$
of the bi-scalar variable $u$. These are given by
\footnote{$\de_\mu=\frac{\de}{\de z^\mu}\,, 
\de_\mup=\frac{\de}{\de w^\mup}.$}
\beq
  \partial_{\mu} \partial_{\nu'}u= -
  {1 \over z_0w_0} \big [ \delta_{\mu \nu'} +
  {1 \over w_0} (z-w)_{\mu} \delta_{\nu'0} +
  {1 \over z_0} (w-z)_{\nu'}\delta_{\mu 0} -
  u \delta_{\mu 0} \delta_{\nu'0} ],
\eeq 
and $\partial_{\mu}u \partial_{\nu'}u$ with
\begin{eqnarray}
  \partial_{\mu} u &=& {1 \over z_0}
     [(z-w)_{\mu} / w_0 - u \delta_{\mu 0}], \\
   \partial_{\nu'} u &=& {1 \over w_0}
     [(w-z)_{\nu'} / z_0 - u \delta_{\nu' 0}] \, .\nonumber
\end{eqnarray}
We will need certain properties of derivatives of $u$,
most of which were derived in \cite{dhfgauge} and which we list here.
\begin{eqnarray}
&&   \Box u= D^{\mu} \partial_{\mu} u = (d+1)(1+u),  
\label{2.9a} \\
&&      D^{\mu} u \ \partial_{\mu} u = u (2+u),   \label{2.9b} \\
&&      D_{\mu} \partial_{\nu} u = g_{\mu \nu}(1+u),   \label{2.9c}
\\
&&      (D^{\mu}u)\ (D_{\mu} \partial_{\nu}\partial_{\nu'}u)
          = \partial_{\nu}u \partial_{\nu'}u,   \label{2.9d} \\
&&      (D^\mu u) \ (\dmu \dnup u)  = (1+u) \dnup u, \label{2.9e}
\\
&&      (D^\mu \dmup u) \ (\dmu \dnup u)  = g_{\mu ' \nu '} +
\dmup u \dnup u,  \label{2.9f} \\
&&      D_\mu \dnu \dnup u  = g_{\mu \nu} \dnup u.  \label{2.9g}
\end{eqnarray}

\section{Antisymmetric Tensor}
\setcounter{equation}{0}
The equation of motion for an anti-symmetric tensor field in 
$AdS_{d+1}$ is
\beq
\frac{1}{2}D^\mu \partial_{[\mu} A_{\nu \alpha]} 
- m^2 A_{\nu \alpha} = J_{\nu \alpha}.
\label{max}
\eeq
The covariant derivative is with respect to the $AdS$ metric. 
 The Maxwell operator is normalized such that 
$\Box A_{\mu \nu}$ appears 
with coefficient 1 in the equation of motion. 
We look for solutions
of the form
\beq
A_{\mu \nu}=\int d^{d+1}w \sqrt{g} G_{\mu \nu}{}^{\mu^\prime \nu^\prime}(z,w)
J_{\mu^\prime \nu^\prime}(w),
\label{1}
\eeq
with bi-tensor propagator $G_{\mu \nu}{}^{\mu^\prime \nu^\prime}(z,w)$. Using this
expression for $A_{\mu \nu}$ in Eq(\ref{max}), we obtain an equation for
the propagator $G_{\mu \nu}{}^{\mu^\prime \nu^\prime}(z,w)$\footnote
{[...] denotes anti-symmetrization with strength 1}: 
\begin{equation}
\frac{1}{2}D^\mu \partial_{[u}G_{\nu \alpha]}{}^{\nu^\prime \alpha^\prime}(z,w)
-m^2 G_{\nu \alpha}{}^{\nu^\prime \alpha^\prime}(z,w) = - \delta(z,w)
(g_{\nu}{}^{ \nu^\prime} g_{\alpha}{}^{ \alpha^\prime}-
g_{\alpha}{}^{ \nu^\prime} g_{\nu}{}^{ \alpha^\prime}).
\label{proca}
\end{equation}
\subsection*{$m=0$}
We will first look at the case $m=0$. In this case, there is a gauge invariance
of the form $A_{\mu \nu} \rightarrow A_{\mu \nu} + \partial_\mu \Lambda_\nu
- \partial_\nu \Lambda_\mu$ which implies that the current $J^{\mu \nu}$ is
conserved. The equation for 
$G_{\mu \nu}{}^{\mu^\prime \nu^\prime}(z,w)$ is:
\begin{equation}
\frac{1}{2}D^\mu \partial_{[\mu}G_{\nu \alpha]}{}^{\nu^\prime \alpha^\prime}(z,w)
 = - \delta(z,w)
(g_{\nu}{}^{ \nu^\prime} g_{\alpha}{}^{ \alpha^\prime}-g_{\alpha}{}^{ \nu^\prime} g_{\nu}{}^{ \alpha^\prime})
+ \partial^{\nu^\prime} \Lambda_{\nu \alpha}{}^{\alpha^\prime} - 
\partial^{\alpha^\prime} \Lambda_{\nu \alpha}{}^{\nu^\prime},
\label{eq3}
\end{equation}
where the second term on the right hand side gives zero when integrated
with conserved currents. We will introduce two bi-tensors,
\begin{eqnarray}
T_{\mu \nu}{}^{\mup \nup}& = &(\de_\mu \de^\mup u \de_\nu \de^\nup u- \de_\mu \de^\nup u \de_\nu \de^\mup u), \\
S_{\mu \nu;\mup \nup}& = &(\de_\mu \de^\mup u \de_\nu u \de^\nup u- \de_\mu \de^\nup u  \de_\nu  u \de^\mup u
-\de_\nu \de^\mup u  \de_\mu  u \de^\nup u + \de_\nu \de^\nup u \de_\mu u \de^\mup u).
\end{eqnarray} 
These are the only two bi-tensors which are anti-symmetric under
 $\mu \leftrightarrow \nu$ and $\mup \leftrightarrow \nup$. 
We will choose the following ansatz for 
$G_{\mu \nu}{}^{\mu^\prime \nu^\prime}(z,w)$:
\[
G_{\nu \alpha}{}^{\nu^\prime \alpha^\prime}(z,w) =T_{\nu \al}{}^{\nup \alp}F(u)+
\partial_{\nu} L_{\al}{}^{\nu^\prime \alp}-
\partial_{\al} L_{\nu}{}^{\nu^\prime \alp}.
\]
The second term is a pure gauge and is annihilated by the Maxwell operator:
\begin{eqnarray*}
\partial_{[\mu}G_{\nu \alpha]}{}^{\nu^\prime \alpha^\prime}& =& T_{[\nu \al}{}^{\nup \alp}
\partial_{\mu]}F^\pr(u) \\ & = & (\partial_{[\nu}
\partial^{\nup}u \partial_\al \partial^{\alp}u \partial_{\mu]}u
-\partial_{[\nu}
\partial^{\alp}u \partial_\al \partial^{\nup}u \partial_{\mu]}u) F^\prime(u).
\end{eqnarray*}
In Eq(\ref{eq3}), we need an expression for 
$D^\mu \partial_{[\mu} G_{\nu \alpha]}{}^{\nup \alp}$: 
\begin{eqnarray}
D^\mu \partial_{[\mu} G_{\nu \alpha]}{}^{\nup \alp}& =&
D^\mu( T_{[\nu \al}{}^{\nup \alp}\partial_{\mu]}F^\pr(u)) \nonumber \\
& = &  
(\partial^{\nup}u) \partial_\al \partial^{\alp}u \partial_{\mu]}u F^\prime(u)
+ \partial_{[\nu}
\partial^{\nup}u(D^\mu  \partial_\al \partial^{\alp}u) \partial_{\mu]}u F^\prime(u) \nn\\
& &  
+\partial_{[\nu}
\partial^{\nup}u \partial_\al \partial^{\alp}u (D^\mu \partial_{\mu]}u) F^\prime(u)
+\partial_{[\nu}
\partial^{\nup}u \partial_\al \partial^{\alp}u \partial_{\mu]}uD^\mu u  F^{\prime\prime}(u) \nn\\
& & - \nup \leftrightarrow \alp. \label{main}
\end{eqnarray}
Various terms in Eq(\ref{main}) are simplified using properties of derivatives of $u$
(Eqs \ref{2.9a}-\ref{2.9g}):
\[
(D^\mu \partial_{[\nu}
\partial^{\nup}u) \partial_\al \partial^{\alp}u \partial_{\mu]}u F^\prime(u)
=(\partial^\nup u \partial_\nu \partial^\alp u \partial_\al u
-\partial^\nup u \partial_\alpha \partial^\alp u \partial_\nu u ) (d-1)F^\prime(u),
\]
\[
\partial_{[\nu}
\partial^{\nup}u(D^\mu  \partial_\al \partial^{\alp}u) \partial_{\mu]}u F^\prime(u)
=- (\partial^\alp u \partial_\nu \partial^\nup u \partial_\al u 
-\partial^\alp u \partial_\nu u \partial_\al  \partial^\nup u) (d-1)F^\prime(u),
\]
\[
\partial_{[\nu}
\partial^{\nup}u \partial_\al \partial^{\alp}u (D^\mu \partial_{\mu]}u) F^\prime(u)
= T_{\nu \al}{}^{\nup \alp}
(d-1)(1+u)F^\prime(u),
\]
\[ 
\partial_{[\nu}
\partial^{\nup}u \partial_\al \partial^{\alp}u 
\partial_{\mu]}uD^\mu u  F^{\prime\prime}(u)
= u(2+u)T_{\nu \al}{}^{\nup \alp} F^{\prime \prime} 
 -(1+u)S_{\al \nu}{}^{\alp \nup} F^{\prime \prime}(u).
\]
Collecting all the terms together, 
\begin{eqnarray}
D^{\mu} \partial_{[\mu}G_{\nu \alpha]}{}^{\nu^\prime \alpha^\prime}& = &
 2 \Bigl(u(2+u)F^{\prime \prime}+ (d-1)(1+u)F^\prime \Bigr)
T_{\nu \al}{}^{\nup \alp} \nn \\ & & -2\Bigl((d-1)F^\prime  +   (1+u) F^{\prime \prime}\Bigr)S_{\al \nu}{}^{\alp \nup}.
\label{maxg}
\end{eqnarray}
Using $AdS$ invariance and the fact that $\Lambda_{\nu \al;\alp}$ is anti-symmetric
in $\nu$ and $\al$, $\Lambda_{\nu \al}{}^{\alp}$ has to be of the form:
\begin{equation}
\Lambda_{\nu \al}{}^{\alp}= (\de^\alp \de_\nu u \de_\al u - \de^\alp \de_\al u \de_\nu u) \Lambda(u),
\end{equation}
where $\Lambda$ is a scalar function of $u$. Then, 
\[
\partial^{\nu^\prime} \Lambda_{\nu \alpha}{}^{\alpha^\prime} - 
\partial^{\alpha^\prime} \Lambda_{\nu \alpha}{}^{\nu^\prime}
= -2 \Lambda T_{\nu \al}{}^{\nup \alp}+ \Lambda^\pr S_{\al \nu}{}^{\alp \nup}.
\]
Using these expressions, Eq(\ref{eq3}) becomes (for $z$ and $w$ separate)
\begin{eqnarray*}
&&\Bigl(u(2+u)F^{\prime \prime}+ (d-1)(1+u)F^\prime+2\Lambda\Bigr)
T_{\nu \al}{}^{\nup \alp} \\
&&-\Bigl((d-1)F^\prime + (1+u) F^{\prime \prime}-\Lambda^\pr \Bigr)
S_{\al \nu}{}^{\alp \nup}=0.
\end{eqnarray*}
Setting the scalar coefficients of the two independent tensors to zero, we
obtain
\begin{eqnarray}
u(2+u)F^{\prime \prime} + (d-1)(1+u)F^\prime & = & -2\Lambda, \label{one} \\
(1+u)F^{\prime \prime}+ (d-1)F^\prime & = &  \Lambda^\prime.
\label{easy}
\end{eqnarray}
Eq (\ref{easy}) can be integrated to give\footnote{We consistently ignore
constants of integration since we want our propagators to go to zero
as $u \rightarrow \infty$.}
\begin{equation}
\Lambda=(d-2)F+(1+u)F^\prime,
\end{equation}
which can be substituted into (\ref{one}):
\begin{equation}
u(2+u)F^{\prime \prime}+(d+1)(1+u)F^\prime + 2(d-2)F =0.
\end{equation}
This is just the invariant equation
\beq
(\Box-\mu^2)F(u)=0
\eeq
for the propagator of a scalar field of $\mu^2=-2(d-2)$. A scalar field of mass $m^2$ is
characterized by two possible scale dimensions, 
\beq
\Delta_{\pm}=\frac{d}{2}\pm \frac{1}{2}\sqrt{d^2+4\mu^2}.
\eeq
For our propagator to have the fastest fall off as $u \rightarrow \infty$, 
in the following solution, we will choose $\Delta = \Delta_{+}$.  
\begin{eqnarray}
F(u) & = & \tilde{C}_{\Delta} (2u^{-1})^\Delta F(\Delta, \Delta-\frac{d}{2}+\frac{1}{2}; 2\Delta -d+1;
-2u^{-1}), \label{sol}\\
 \tilde{C}_{\Delta} & = & \frac{\Gamma(\Delta)\Gamma(\Delta-\frac{d}{2}+\frac{1}{2})}
{(4 \pi )^{(d+1)/2} \Gamma(2 \Delta -d +1)}, \nn \\
\Delta & = & \frac{d}{2}+ \sqrt{\frac{d^2}{4}-2(d-2)}, \nn \\
\end{eqnarray}
where $F$ is the standard hypergeometric function $ _2 F_1$. 
The constant $C_\Delta$ is chosen such that as $u \rightarrow 0$, 
$F(u)$ matches on to the flat space case (this is discussed in more
detail in section 6). 
\subsection*{$m\neq 0$}
We will now consider the massive case. In this case, there is no gauge invariance
and the current $J^{\mu \nu}$ is not necessarily conserved.
We will still use the same ansatz, 
\beq
G_{\nu \al}{}^{ \nup \alp}=F(u) T_{\nu \al}{}^{ \nup \alp} + (\de_\nu L_{\al}{}^{ \nup \alp}- \de_\al L_{\nu}{}^{ \nup \alp}).
\label{puregauge}
\eeq
Using $AdS$ invariance, we can write $L_{\al; \nup \alp}$ as
\beq
L_{\al}{}^{ \nup \alp}=(\de_\al \de^\alp u \de^\nup u - \de_\al \de^\nup u \de^\alp u)L(u),
\eeq
where $L(u)$ is a scalar function. 
$D^{\mu} \partial_{[\mu}G_{\nu \alpha];\nu^\prime \alpha^\prime}$ is still given 
by Eq(\ref{maxg}) since the Maxwell operator annihilates the second term in 
Eq(\ref{puregauge}). This term is
\beq
\de_\nu L_{\al}{}^{ \nup \alp}- \de_\al L_{\nu}{}^{ \nup \alp}=2L T_{\al \nu}{}^{\alp \nup}+ L^\pr S_{\al \nu}{}^{\alp \nup}.
\eeq
Eq(\ref{proca}) then becomes (for $z$ and $w$ separate)
\begin{eqnarray*}
  \Bigl(u(2+u)F^{\prime \prime}+ (d-1)(1+u)F^\prime \Bigr)
T_{\nu \al;\nup \alp}   
-\Bigl((d-1)F^\prime  +   (1+u) F^{\prime \prime}\Bigr)S_{\al \nu;\alp \nup}& & \\-m^2 F 
T_{\nu \al;\nup \alp}  -m^2 (2L T_{\al \nu;\alp \nup} +   L^\pr S_{\al \nu;\alp \nup}) = 0. & & 
\end{eqnarray*}
Setting the coefficient of independent tensors to zero, we get a 
system of two coupled differential equations:
\beq
 u(2+u)F^{\prime \prime}+ (d-1)(1+u)F^\prime-m^2 F -2 m^2 L =0,
\label{one2}
\eeq
\beq
-(d-1)F^\prime  -  (1+u) F^{\prime \prime}-m^2 L^\pr =0.
\label{easy2}
\eeq 
Eq(\ref{easy2}) can be readily integrated to give
\beq
L= -\frac{1}{m^2} (( d-2)F + (1+u)F^\pr).
\eeq
Substituting in Eq(\ref{one2}), we find an uncoupled differential
equation for $F(u)$
\beq
u(2+u)F^{\prime \prime}+ (d+1)(1+u)F^\prime+(-m^2+2(d-2)) F  =0.
\eeq
This is again the invariant equation
\beq
(\Box -\mu^2)F(u)=0
\eeq
for the propagator of a scalar field of $\mu^2=(m^2-2(d-2))$. 
\section{\boldmath $p$-forms}
\setcounter{equation}{0}
The preceding calculation of the propagator of an anti-symmetric tensor can be 
generalized to $p$-form propagators.  
The equations of motion for a $p$-form field is
\begin{equation}
\frac{1}{p!}D^\al \de_{[\al} A_{\mo \mt \dots \msp]} -m^2 A_{\mo \mt \dots \msp} =0.
\eeq
We will assume a solution of the form
\[
A_{\mo \mt \dots \msp}=\int d^{d+1}w \sqrt{g} G_{\mo \mt \dots \msp}{}^{ \mop \mtp \dots \mpp}(z,w)
J_{\mop \mtp \dots \mpp}(w),
\]
with bi-tensor propagator $G_{\mo \mt \dots \msp}{}^{ \mop \mtp \dots \mpp}(z,w)$. The propagator
satisfies the equation:
\begin{equation}
\frac{1}{p!}D^\al \partial_{[\al}G_{\mo \mt \dots \msp]}{}^{\mop \mtp \dots \mpp}(z,w)
-m^2 G_{\mo \mt \dots \msp}{}^{ \mop \mtp \dots \mpp}(z,w) = - \delta(z,w)
g_{[\mo}{}^{\mop} g_{\mt}{}^{\mtp}\dots g_{\msp]}{}^{ \mpp}
\label{procap}
\end{equation}
There are two independent
tensors which have the right anti-symmetry property under exchange
of various indices. These are
\begin{eqnarray}
T_{\mo \mt \dots \msp}{}^{ \mop \mtp \dots \mpp} & =& 
 \de_{[\mo} \de^{\mop} u \de_\mt \de^\mtp u \dots \de_{\msp]} \de^{\mtp}u, \\
S_{\mo \mt \dots \msp}{}^{ \mop \mtp \dots \mpp} & =& 
 \de_{[\mo} u\de^{[\mop} u \de_\mt \de^\mtp u \dots \de_{\msp]} \de^{\mtp]}u.
\end{eqnarray}
Generalizing the ansatz of the last section  for $G_{\mo \mt \dots \msp}{}^{\mop \mtp \dots \mpp}(z,w)$, we try a solution of the form
\beq
G_{\mo \mt \dots \msp}{}^{ \mop \mtp \dots \mpp}=F(u) T_{\mo \mt \dots \msp}{}^{ \mop \mtp \dots \mpp}
+\de_{[\mo}L_{\mt \dots \msp]}{}^{ \mop \mtp \dots \mpp},
\eeq
where 
\beq
L_{\mt \dots \msp}{}^{ \mop \mtp \dots \mpp}=\Bigl(\de^{[\mop} u   \de_\mt \de^\mtp u \dots \de_{\msp} \de^{\mpp]}u\Bigr) L(u).
\eeq
Then,
\beq
\de_{[\mo}L_{\mt \dots \msp]}{}^{ \mop \mtp \dots \mpp}= p L T_{\mo \mt \dots \msp}{}^{ \mop \mtp \dots \mpp}
+L^\pr S_{\mo \mt \dots \msp}{}^{ \mop \mtp \dots \mpp},
\eeq
and
\beq
\de_{[\al} G_{\mo \mt \dots \msp]}{}^{ \mop \mtp \dots \mpp}= F^\pr \de_{[\al} u 
T_{\mo \mt \dots \msp]}{}^{\mop \mtp \dots \mpp}.
\eeq
Notice that the second term in
 $G_{\mo \mt \dots \msp}{}^{ \mop \mtp \dots \mpp}$ 
(of the form $\partial L$ does not contribute). 
Acting with $D^\al$, we get
\begin{eqnarray*}
D^\al 
\de_{[\al} G_{\mo \mt \dots \msp]}{}^{ \mop \mtp \dots \mpp} & = &
F^{\pr \pr} D^\al u  \de_{[\al} u T_{\mo \mt \dots \msp]}{}^{ \mop \mtp \dots \mpp}
 + F^\pr (D^\al \de_{[\al} u) T_{\mo \mt \dots \msp]}{}^{ \mop \mtp \dots \mpp} \\
& + &F^\pr \de_{[\al} u D^\al T_{\mo \mt \dots \msp]}{}^{ \mop \mtp \dots \mpp}.
\end{eqnarray*}
Various terms appearing on the r.h.s of the above equation can be simplified
by using Eq(\ref{2.9a}-\ref{2.9g}):
\[
D^\al u  \de_{[\al} u T_{\mo \mt \dots \msp]}{}^{ \mop \mtp \dots \mpp}=
p!(u(2+u)T_{\mo \mt \dots \msp}{}^{ \mop \mtp \dots \mpp}-(1+u)S_{\mo \mt \dots \msp}{}^{\mop \mtp \dots \mpp}),
\]
\[
(D^\al \de_{[\al} u) T_{\mo \mt \dots \msp]}{}^{\mop \mtp \dots \mpp}=
p!((d+1-p)(1+u)T_{\mo \mt \dots \msp}{}^{ \mop \mtp \dots \mpp}),
\]
\[
\de_{[\al} u D^\al T_{\mo \mt \dots \msp]}{}^{ \mop \mtp \dots \mpp}=
-p! (d+1-p)S_{\mo \mt \dots \msp}{}^{ \mop \mtp \dots \mpp}. 
\]
So (\ref{procap}) becomes (for $z$ and $w$ separate), 
\begin{eqnarray*}
&   & \frac{1}{p!}D^\al \de_{[\al} G_{\mo \mt \dots \msp]}{}^{ \mop \mtp \dots \mpp}-m^2
G_{\mo \mt \dots \msp}{}^{\mop \mtp \dots \mpp}=\\
&&\Bigl( u(2+u) F^{\pr \pr}+(d+1-p)F^\pr-m^2F
-pm^2L\Bigr)
T_{\mo \mt \dots \msp}{}^{ \mop \mtp \dots \mpp}\\  &&  -\Bigl((1+u)F^{\pr \pr}+(d+1-p)F^\pr)-m^2 L^\pr \Bigr)
S_{\mo \mt \dots \msp}{}^{ \mop \mtp \dots \mpp} =0,
\end{eqnarray*}
which implies
\beq
 u(2+u) F^{\pr \pr}+(d+1-p)F^\pr -m^2F
-pm^2L=0,
\label{eq1}
\eeq
\beq
(1+u)F^{\pr \pr}+(d+1-p)F^\pr+m^2 L^\pr=0.
\label{eq}
\eeq
Eq(\ref{eq}) can be integrated to give
\beq
L=-\frac{1}{m^2}\Bigl((d-p)F+(1+u)F^\pr \Bigr),
\eeq
which can then be substituted in Eq(\ref{eq1}):
\beq
u(2+u) F^{\pr \pr}+(d+1)F^\pr+\Bigl(-m^2 +p(d-p)\Bigr)F =0.
\eeq
This is the equation for the propagator of the scalar field of $\mu^2=m^2-p(d-p)$.
The solution to this equation is given by Eq(\ref{sol}) with 
\[
\Delta=\frac{d}{2}+\frac{1}{2}\sqrt{d^2+4m^2-4p(d-p)}.
\]
This result agrees with \cite{iosifp} for the case $m=0$.

\section{Massive Symmetric Tensor}
\setcounter{equation}{0}
We will begin by reviewing the equations of motion for the massless graviton.
The propagator for the graviton was obtained in \cite{dfmmr}. The gravitational
action is
\beq
 I_g = {1\over2\kappa^2} \int d^{d+1} z \sqrt g (R - \Lambda)+S_m,
\eeq
where $S_m$ is the matter action. 
The equations of motion are
\beq
 R_{\mu \nu} - {1\over 2} g_{\mu \nu} (R-\Lambda)  = T_{\mu \nu}.
\eeq
$T_{\mu \nu}$ is the stress-energy tensor. 
For $\Lambda=-d(d-1)$ and $T_{\mu \nu}=0$, we obtain Euclidean $AdS$ space
with $R=-d(d+1)$. In the presence of a matter source, the metric will no longer
be the $AdS$ metric $g_{\mu \nu}$. We will denote the fluctuations about $g_{\mu \nu}$
by $h_{\mu \nu}$. An equivalent form of the equation is
\beq
R_{\mu\nu} + d g_{\mu\nu} = \tilde T _{\mu \nu}
           \equiv T_{\mu \nu} - {1 \over d-1} g_{\mu \nu}
T_\sigma{}^\sigma. 
\eeq
The linearized equations of motion for $h_{\mu \nu}$ are
\beq
 -  D^\sigma D_\sigma h_{\mu \nu} -  D_\mu D_\nu h_\sigma
{}^\sigma +  D_\mu D^\sigma h_{ \sigma \nu} +  D_\nu
D^\sigma h_{\mu \sigma} -2(h_{\mu \nu} - g_{\mu \nu}
h_\sigma {}^\sigma) = 2 \tilde T_{\mu \nu}.
\eeq
All covariant derivatives and contractions are with respect to $g_{\mu \nu}$.

For the case of massive symmetric tensor, $S_{\mu \nu}$, a consistent action
with coupling to gravity is not known. For example,
consider a Kaluza-Klein reduction of a theory 
of gravity in $D$ dimensional space-time to $D-1$ dimensional space time. We 
get an infinite tower of massive symmetric tensor fields in $D-1$ dimensions. 
It was shown in \cite{duff} that it is impossible to consistently truncate
this theory to a finite number of symmetric tensor fields. However, the quadratic
part of the action (which is what we need for calculation of the propagator) was
given in \cite{buchbinder} \cite{Buchbinder2}.
\begin{eqnarray*}
I&=&\int d^{d+1}z \sqrt{g}\Bigl( \frac{1}{4}D_\mu S D^\mu S -\frac{1}{4}
D_\mu S_{\nu \al}D^{\mu}S^{\nu \al}-\frac{1}{2}D^{\mu}S_{\mu \nu}D^{\nu}S
+ \frac{1}{2}D_\mu S_{\nu \al}D^{\al}S^{\nu \mu}\nn \\
& +& \frac{1}{2}S_{\mu \nu}S^{\mu \nu}-\frac{d-1}{2} S^2 -\frac{m^2}{4}S_{\mu \nu}
S^{\mu \nu}+\frac{m^2}{4}S^2- S_{\mu \nu}J^{\mu \nu}\Bigr),
\end{eqnarray*}
where $S=g_{\mu \nu}S^{\mu \nu}$. 
From this action, we can derive the following equation of motion:
\begin{eqnarray*}
&-&\frac{1}{2}D^\sigma D_{\sigma} S_{\mu \nu} - \frac{1}{2} D_\mu D_\nu S
+\frac{1}{2} D_\mu D^\sigma S_{\sigma \nu}+\frac{1}{2} D_\nu D^\sigma 
S_{\sigma \mu}
-S_{\mu \nu}+\frac{1}{2} g_{\mu \nu}D^\sigma D_\sigma S \nn \\
&-& \frac{1}{2}g_{\mu \nu}D_\mu D_\sigma S^{\sigma \mu}- \frac{d-1}{2}g_{\mu \nu}S-\frac{m^2}{2}S_{\mu \nu}+\frac{m^2}{2} g_{\mu \nu} S=
J_{\mu \nu}.
\end{eqnarray*}
This can be converted to the following equivalent form
\beq
 -  D^\sigma D_\sigma S_{\mu \nu} -  D_\mu D_\nu S_\sigma
{}^\sigma +  D_\mu D^\sigma S_{ \sigma \nu} +  D_\nu
D^\sigma S_{\mu \sigma} -2(S_{\mu \nu} - g_{\mu \nu}
S_\sigma {}^\sigma)+m_1^2 S_{\mu \nu}+m_2^2 g_{\mu \nu}S_\sigma^\sigma = 2 \tilde J_{\mu \nu},
\label{linearized}
\eeq
where $m_2^2=\frac{m_1^2}{d-1}=\frac{m^2}{d-1}$ and $\tilde J_{\mu \nu}= J_{\mu \nu}-
\frac{1}{d-1}J_\sigma {}^\sigma$. We work with arbitrary $m_1$ and $m_2$ and
only use the relation $m_2^2=\frac{m_1^2}{d-1}$ towards the end
of the calculation. 

We look for solutions of the form
\beq
S_{\mu\nu}(z)=\int d^{d+1}w \sqrt g G_{\mu \nu;\mu ' \nu'}(z,w)
  {J}^{\mu'\nu'}(w),
\eeq
where $G_{\mu \nu;\mu ' \nu'}(z,w) $is the bi-tensor propagator for the massive
symmetric tensor. Inserting this expression in Eq(\ref{linearized}), we obtain
the equation
\begin{eqnarray}
\label{graviton}
 &-&  D^\sigma D_\sigma G_{\mu \nu;\mu ' \nu '}
  -  D_\mu D_\nu G_{\sigma} {}^\sigma {}_{\mu ' \nu '}
  +  D_\mu D^\sigma G_{ \sigma \nu; \mu ' \nu '} \\
  &+&  D_\nu D^\sigma G_{\mu \sigma ; \mu ' \nu '}
-2(G_{\mu \nu; \mu ' \nu '} - g_{\mu \nu} G_\sigma {}^\sigma
{}_{;\mu ' \nu'})+m_1^2 G_{\mu \nu;\mu ' \nu '}+m_2^2 g_{\mu \nu}
G_{\sigma}{}^ \sigma{}_{;\mu ' \nu '} \nonumber\\  
 &=&
 \Bigl(g_{\mu \mu'}g_{\nu \nu'} +g_{\mu \nu'}g_{\nu \mu'} -  {2\over d-1}
g_{\mu\nu}g_{\mu' \nu'}\Bigr)
\delta(z,w). \nn
\end{eqnarray}
The next step is to write $G_{\mu \nu;\mu ' \nu'}(z,w) $ in terms of
invariant bi-tensors in $AdS_{d+1}$ and scalar functions.  
We will use the following five bi-tensors defined in \cite{dfmmr}:
\begin{eqnarray}
 T^{(1)} _{\mu \nu;\mu' \nu'} &=&
 g_{\mu \nu} \ g_{\mu ' \nu'},
\label{T's} \\ T^{(2)} _{\mu \nu;\mu' \nu'} & = &
 \dmu u \ \dnu u \ \dmup u \ \dnup u,
\nonumber \\ T^{(3)} _{\mu \nu;\mu' \nu'} & = &
 \dmu \dmup u \ \dnu \dnup u + \dmu \dnup u \ \dnu \dmup u,
 \nonumber\\ T^{(4)} _{\mu \nu;\mu' \nu'} & = &
 g_{\mu \nu} \ \dmup u \ \dnup u + g_{\mu ' \nu'} \ \dmu u \
\dnu u,
 \nonumber\\ T^{(5)} _{\mu \nu;\mu' \nu'} & = &
 \dmu \dmup u \ \dnu u \dnup u + \dnu \dmup u \ \dmu u \dnup
u
 + (\mu' \leftrightarrow \nu').
\nonumber
\end{eqnarray}
Our ansatz for the propagator $G_{\mu \nu;\mu ' \nu'}(z,w) $ is
\begin{eqnarray*}
G_{\mu \nu ;\mu' \nu'}& =& (\partial_\mu \partial_{\mu'} u\,\partial_\nu \partial_{\nu'} u+\partial_\mu \partial_{\nu'} u\,\partial_\nu \partial_{\mu'} u) \,G(u) +g_{\mu \nu} g_{\mu' \nu'} \,H(u) \\
& & +D_\mu L_{\nu;\mup \nup}+D_\nu L_{\mu; \mup \nup} +
D_\mup \Lambda_{\mu \nu;\nup}+D_\nup \Lambda_{\mu \nu;\mup},
\end{eqnarray*}
where
\begin{eqnarray}
\Lambda_{\nu;\mup \nup}&=&g_{\mup \nup} \dnu u A(u) + \dmup u \dnup u \dnu u C(u)
+(\dmup \dnu u \dnup u + \dnup \dnu u \dmup u) B(u) \\
\Lambda_{\mu \nu;\nup} & = & g_{\mu \nu} \de_{\nup} u A(u) + \dmu u \dnu u \dnup u C(u)
+(\dmu \dnup u \dnu u + \dnu \dnup u \dmu u) B(u),
\end{eqnarray}
and
\begin{eqnarray}
D_\mu L_{\nu;\mup \nup} + D_\nu L_{\mu;\mup \nup} & = & 2(1+u)AT^{(1)} + 
g_{\mup \nup} \dmu u \dnu u 2A' +2C'T^{(2)} +2BT^{(3)} + C T^{(5)} \nonumber \\
& + & g_{\mu \nu} \dmup u \dnup u (2(1+u)C+4B) +B'T^{(5)},
\end{eqnarray}
\begin{eqnarray}
D_\mup \Lambda_{\mu \nu;\nup} + D_\nup \Lambda_{\mu \nu;\mup} & = & 2(1+u)AT^{(1)} + 
g_{\mu \nu} \dmup u \dnup u 2A'+2C'T^{(2)} +2BT^{(3)} + C T^{(5)} \nonumber \\
& + & g_{\mup \nup} \dmu u \dnu u (2(1+u)C+4B)  +B'T^{(5)}.
\end{eqnarray}
This ansatz is invariant under exchange $z\leftrightarrow w$ (exchange of
primed and unprimed indices). 
We note that
\begin{eqnarray*}
D_\mup \Lambda_{\mu \nu;\nup} + D_\nup \Lambda_{\mu \nu;\mup} &=&
D_\mu L_{\nu;\mup \nup} + D_\nu L_{\mu;\mup \nup}+
\Bigl(-2A^\pr+2(1+u)C+4B \Bigr)g_{\mup \nup}\de_\mu u \de_\nu u \\
& - &\Bigl(-2A^\pr+2(1+u)C+4B \Bigr)g_{\mu \nu}\de_\mup u \de_\nup u.
\end{eqnarray*} 
The term $D_\mu L_{\nu;\mup \nup} + D_\nu L_{\mu;\mup \nup}$
just corresponds to diffeomorphisms in $z$ and is annihilated by the
``wave operator''. Unlike the massless graviton case, 
all five functions $G(u)$, $H(u)$, $A(u)$, $B(u)$ and $C(u)$ are physical.
We now find the differential equations for these five functions. 
Using this ansatz in (\ref{graviton}) and using relations in 
Eq(\ref{2.9a}-\ref{2.9g}),
 it is simple, but tedious to derive
the following equation:
\begin{eqnarray}
 & - &  D^\sigma D_\sigma G_{\mu \nu;\mu ' \nu '}
  -  D_\mu D_\nu G_{\sigma} {}^\sigma {}_{\mu ' \nu '}
  +  D_\mu D^\sigma G_{ \sigma \nu; \mu ' \nu '}
  +  D_\nu D^\sigma G_{\mu \sigma ; \mu ' \nu '}
\\  & - &2(G_{\mu \nu; \mu ' \nu '} - g_{\mu \nu}
G_\sigma {}^\sigma {}_{;\mu ' \nu'}) +m_1^2 G_{\mu \nu;\mu ' \nu '}
+m_2^2 g_{\mu \nu} G_{\sigma}{}^ \sigma{}_{;\mu ' \nu '}
\nonumber\\  & = &
\ \ T^{(1)} \Bigl[-u(2+u)H'' -2d (1+u) H' +2dH +4G
-2(1+u)G' +4(1+u)m_1^2 A \nonumber \\ & +&m_1^2 H
+ m_2^2 (d+1)H+4m_2^2(d+1)(1+u)A +2m_2^2 u(2+u) A^\pr
+8m_2^2B\nonumber \\ &+&2m_2^2G+2m_2^2u(2+u)(1+u)C+4m_2^2u(2+u)B\nonumber \\
& + & ((d+1)u^2+(2d+2)u+d)(-4A^\pr +4(1+u)C+8B)\nonumber \\ &+&
u(2+u)(1+u)(-2A^{\pr \pr}+2(1+u)C^\pr+2C+4B^\pr)+4(1+u)C+8B-4A^\pr
\Bigr]  \nonumber \\& +& 
 T^{(2)} \Bigl [ - 2 G'' + 4 m_1^2 C'-(d-1)(-2(1+u)C^{\pr \pr}-4C^\pr-4B'' +2A''') \Bigr ] \nonumber \\
& + &  T^{(3)} \Bigl [-u(2+u) G'' - (d-1) (1+u) G' + 2(d-1) G +4m_1^2 B
+m_1^2G \nonumber \\ &-&(d-1)(-2(1+u)C-4B+2A')\Bigr ]
\nonumber\\ & + & g_{\mu \nu} \dmup u \dnup u \Bigl [2(1+u) G'
+ 4(d+1) G +4m_1^2 B + 2(1+u)m_1^2C+2m_2^2 (1+u)(d+1)C \nonumber \\
&+&2(m_1^2+m_2^2(d+1)) A^\pr   
 + 8m_2^2 (1+u) C 
 +  8m_2^2 (1+u)B^\pr\nonumber \\ &+&4m_2^2 u(2+u)C^\pr+8m_2^2 B+2m_2^2G
+4m_2^2(d+1)B \nonumber \\
&-&2(d+2)(1+u)(-2(1+u)C'-2C-4B'+2A'')\nonumber \\ 
&-&u(2+u)(-2(1+u)C''-4C'-4B''+2A''') \nonumber \\
&-& 2(d+1)(-2(1+u)C-4B+2A')\Bigr] \nonumber \\ &+&  g_{\mu ' \nu'} \dmu u \dnu u \Bigl [-2 G'' -
(d-1) H'' +2m_1^2 A^\pr +4m_1^2 B +2m_1^2 (1+u) C \nonumber \\ &+&2(d-1)(-2A'+2(1+u)C+4B)
\nonumber \\ & + & (1+u)(d-1)(-2A''+2(1+u)C'+2C+4B')\Bigr ]
\nonumber\\ & + & T^{(5)} \Bigl [(1+u)G'' +(d-1)G' +2m_1^2 B'+2m_1^2C 
\nonumber \\ &+&(d-1)(2(1+u)C'+2C+4B'-2A'')\Bigr]\, .
\nonumber
\end{eqnarray}
We have dropped indices from $T^{(i)}$ for clarity. 
We define
\beq
f \equiv -2A^\pr+2(1+u)C+4B.
\label{defnf}
\eeq
For $z$ and $w$ separate, we can set the scalar coefficient 
of each independent tensor to 0.
We then obtain the following system of six differential equations: 
\begin{eqnarray}
&-&u(2+u)H'' -2d (1+u) H' +2dH +4G
-2(1+u)G' +4(1+u)m_1^2 A \nonumber \\ & +&m_1^2 H
+ m_2^2 (d+1)H+4m_2^2(d+1)(1+u)A +2m_2^2 u(2+u) A^\pr
+8m_2^2B\nonumber \\ &+&2m_2^2G+2m_2^2u(2+u)(1+u)C+4m_2^2u(2+u)B\nonumber \\
& + & 2((d+1)u^2+(2d+2)u+d)f+
u(2+u)(1+u)f^\pr+2f  =0, \label{consistency}\\
&& \nonumber\\ 
& -&   2 G'' + 4 m_1^2C'+(d-1)f^{\pr \pr} =0, \label{candg}
 \\&&\nonumber \\
&- &u(2+u) G'' - (d-1) (1+u) G' + 2(d-1) G +4m_1^2 B
+m_1^2G +(d-1)f =0, \label{forg}\\ && \nonumber \\
& & 2(1+u) G'
+ 4(d+1) G +4m_1^2 B + 2(1+u)m_1^2C+2m_2^2 (1+u)(d+1)C \nonumber \\
&+&2(m_1^2+m_2^2(d+1)) A^\pr   
 + 8m_2^2 (1+u) C 
 +  8m_2^2 (1+u)B^\pr +4m_2^2 u(2+u)C^\pr+8m_2^2 B \nn \\ &+&2m_2^2G
+4m_2^2(d+1)B
+2(d+2)(1+u)f^\pr
+u(2+u)f^{\pr \pr}
+ 2(d+1)f=0, \label{69} \\ & & \nonumber \\
&-&2 G'' -
(d-1) H'' +2m_1^2 A^\pr +4m_1^2 B +2m_1^2 (1+u) C +2(d-1)f
\nonumber \\ & +&  (1+u)(d-1)f^\pr=0, \label{forh} \\&& \nonumber \\
& & (1+u)G'' +(d-1)G' +2m_1^2 B'+2m_1^2C 
+(d-1)f^\pr
 =0. \label{forb}
\end{eqnarray}
This is an over determined system since we have six equations for five functions which
can be traced to the fact that Eq(\ref{linearized}) does not possess the symmetry
under exchange of $z$ and $w$ which the propagator itself possesses. 
So we will have to check that one of the above equations is redundant.    

Eq(\ref{candg}) can be integrated to give
\beq
(d-1)f^\pr=2G^\pr-4m_1^2 C,
\label{eqforf}
\eeq
In this and all that follows, we will consistently drop integration constants
since we want our solution to approach $0$ as $u \rightarrow \infty$. 
We can integrate Eq(\ref{eqforf}) to obtain
\beq
(d-1)f=2G-4m_1^2 D,
\label{ab}
\eeq
where 
\beq
C=D^\pr.
\label{cd}
\eeq 
Eq(\ref{ab})
can then be substituted in Eq(\ref{forb}) to obtain
\beq
B-D=-\frac{1}{2m_1^2}\Bigl((1+u)G'+dG \Bigr).
\label{bminusd}
\eeq
Eq(\ref{forg}), can be brought into the following form using Eq(\ref{ab}) and 
Eq(\ref{bminusd}):
\beq
u(2+u)G^{\pr \pr}(u)+(d+1)(1+u)G^\pr(u)-m_1^2 G(u)=0.
\label{eqforg}
\eeq
This is an uncoupled differential equation for $G(u)$.
In fact, this is just the equation for the scalar propagator with $m^2=m_1^2$. 
The solution is 
\begin{eqnarray}
G(u) & = & \tilde{C}_{\Delta} 
(2u^{-1})^\Delta F(\Delta, \Delta-\frac{d}{2}+\frac{1}{2}; 2\Delta -d+1;
-2u^{-1}), \label{normg}\\
 \tilde{C}_{\Delta} & = & \frac{\Gamma(\Delta)\Gamma(\Delta-\frac{d}{2}+\frac{1}{2})}
{(4 \pi )^{(d+1)/2} \Gamma(2 \Delta -d +1)},
\end{eqnarray}
where
\beq
\Delta=\frac{d}{2}+ \frac{1}{2}\sqrt{d^2+4m_1^2}.
\eeq
We now look at Eq(\ref{69}) and
use Eq(\ref{eqforf}), Eq(\ref{ab}) and Eq(\ref{eqforg}) to bring it to the
following form:
\begin{eqnarray*}
& & \Bigl(m_1^2-(d-1)m_2^2\Bigr) u(2+u) D'' +\Bigl(m_1^2-(d-1)m_2^2 \Bigr)(d+5) (1+u)D' \nn \\
&+&\Bigl(m_1^4-4m_1^2+m_1^2m_2^2(d+1)+2m_2^2(d^2+d-2)\Bigr)D \nn \\ 
 &=& -\frac{m_2^2}{m_1^2}(d-1)G''  -\Bigl(-2+ \frac{m_2^2}{m_1^2}(d^2+d-2)\Bigr)(1+u)G'
\nn \\ &-&\Bigl(-2d+ \frac{m_2^2}{m_1^2}d(d^2+d-2)+m_2^2d\Bigr)G.
\label{eqd}
\end{eqnarray*}
We notice that for $m_1=(d-1)m_2^2$  Eq(\ref{eqd}) becomes an algebraic
equation\footnote{This relation between $m_1$ and $m_2$ was discussed 
after Eq(\ref{linearized}).}. 
In this case\footnote{Actually, the expression for $D$ in 
Eq(\ref{d}) is a solution to Eq(\ref{eqd}) for any $m_1$ and $m_2$. It
is the unique solution for $m_1=(d-1)m_2^2$.}  ,
\beq
D=\frac{1}{2d(d-1+m_1^2)m_1^2}\Bigl((d-1)G''+d(d-1)(1+u)G'+(d^3-d^2+m_1^2d)G\Bigr).
\label{d}
\eeq 
From Eq(\ref{cd}), we can get an expression for $C(u)$:
\beq
C=\frac{1}{2d(d-1+m_1^2)m_1^2}\Bigl((d-1)G'''+d(d-1)(1+u)G''+(d^3-d+m_1^2d)G'\Bigr),
\eeq
and from Eq(\ref{bminusd}), we obtain
\beq
B=\frac{1}{2dm_1^2(d-1+m_1^2)}\Bigl((d-1)G''-m_1^2d (1+u)G'-m_1^2d(d-1)G\Bigr).
\eeq
It will be useful to obtain an expression for $f$ from Eq(\ref{ab}):
\beq
f=-\frac{2}{d(d-1+m_1^2)}\Bigl(G''+d(1+u)G'+d(d-1) G\Bigr).
\eeq
Using the definition of $f$ (Eq(\ref{defnf})) and expressions for $B$ and $D$, we obtain
\begin{eqnarray}
A'&=& \frac{1}{2m_1^2d(d-1+m_1^2)}\Bigl((d-1)(1+u)G'''+(d^2+d-2+2m_1^2)G''\nonumber \\
&& -m_1^2d(1+u)G'
+m_1^2 d(d-1)G\Bigr).
\end{eqnarray}
which can be integrated to give
\begin{eqnarray}
A&=& \frac{1}{2m_1^2d(d-1+m_1^2)}\Bigl((d-1)(1+u)G''+(d^2-1+2m_1^2)G'\nonumber \\
&& +m_1^2d(1+u)G + m_1^2 d(d-2) \int G \Bigr).
\end{eqnarray}
We will integrate this once more since the resulting expression for $\int A$
will be used in calculation of $H$. 
\begin{eqnarray}
\int A&=& \frac{1}{2m_1^2d(d-1+m_1^2)}\Bigl((d-1)(1+u)G'+(d^2-d+2m_1^2)G'\nonumber \\
&& +m_1^2d(1+u)\int G +m_1^2 d(d-3)\int \int G \Bigr).
\end{eqnarray}
Eq(\ref{forh}) can be brought into the form:
\beq
(d-1)H=-2G+2m_1^2\int A-2m_1^2(1+u)\int D -2(d-2) \int \int G,
\eeq
from which it follows that
\beq
H=-\frac{2}{d(d-1+m_1^2)}\Bigl((2d^2+m_1^2d-4d) \int \int G + d(1+u)\int G + (d+m_1^2)G\Bigr).
\label{eqforh}
\eeq 

We still need to check that Eq(\ref{consistency}) is redundant. Using 
Eqs(\ref{eqforf},\ref{bminusd},\ref{eqforh},\ref{cd}), 
we can rewrite Eq(\ref{consistency})
in terms of $G(u)$ and $D(u)$ and their derivatives. Then, using
Eq(\ref{eqforg}) and Eq(\ref{eqd}) and integrated versions of these
equations, it is tedious but straight forward to show that the left
hand side is identically zero. 

\section{\boldmath $u \rightarrow 0$}
\setcounter{equation}{0}
\setcounter{footnote}{0}
In this section, we will check the $u \rightarrow 0$ limit of the
propagators for the massive vector and massive symmetric tensor
in $AdS_{d+1}$. We have set the $AdS$ scale $R$ (the radius
of curvature) to unity. Had not set $R$ to unity and taken
 the limit $R \rightarrow \infty$ in our propagators, we should
naively recover the flat space results. This is almost true;
since $m$ in the $AdS$ space is measured in units of $\frac{1}{R}$, 
to match to the flat space propagator of a non-zero mass, we also have
to take $m \rightarrow \infty$. So if we take the $u \rightarrow 0$
and $m \rightarrow \infty$ the propagators should approach the
short distance limit of the corresponding propagators in flat space. 

For comparison with the flat space propagators, it is convenient
to re-write the $AdS_{d+1}$ propagators in terms of $n_\mu$ and $n_\mup$ where
\[
n_\mu (z,w)=D_\mu \mu(z,w), n_\mup(z,w)=D_{\mup}\mu(z,w),
\]
$\mu(z,w)$ is the geodesic distance between $z$ and $w$ and $g_\mu{}^{\nup}$ is
the parallel transporter from $z$ to $w$. We will also need the following
dictionary \cite{dfmmr}
\begin{eqnarray}
u & =& \cosh(\mu)-1, \nn \\
n_\mu & = & \frac{\de_\mu u}{\sqrt{u(u+2)}}, \nn \\
g_{\mu \nup} & = & \de_{\mu} \de_{\nup}u + \frac{\de_\mu u\de_\nup u }{u+2}. 
\label{dictionary}
\end{eqnarray}
\subsection{Massive vector}
For the massive vector, the propagator we obtained in $AdS_{d+1}$
is (specializing the case of $p$-forms in section 3 to $p=1$):
\beq
G_{\mu \mup}(z,w)= (F+L) \de_{\mu} \de_{\mup} u + L^\pr \de_\mu u \de_\mup u.
\label{eq539}
\eeq
In terms of $n_\mu$ and $n_\mup$, we can re-write $G_{\mu \mup}(z,w)$ as
\[
G_{\mu \mup}(z,w)=(F+L)g_{\mu \mup}-\Bigl(u(2+u)L^\pr -u(F+L) \Bigr) n_\mu n_\mup.
\]
As $u\rightarrow 0$, 
\beq
G_{\mu \mup}(z,w)= -\frac{1}{m^2}{\cal F}^\pr g_{\mu \mup}+\frac{2}{m^2}u
{\cal F}^{\pr \pr}n_\mu n_\mup.   
\label{sdlvector}
\eeq
where ${\cal F}$
is just the leading term of $F$ as $u\rightarrow 0$\footnote{${\cal F} \sim \frac{1}{u^{\frac{d-1}{2}}}$}.

In flat space, the propagator for the massive vector is given by
\footnote{$N^\pr(r^2)$ denotes $\frac{d}{d(r^2)}N(r^2)$} 
\begin{eqnarray*}
G^{flat}_{\mu \mup}&=&(g_{\mu \mup}+\de_\mu \de_\mup) N(r^2) \\
& = & (1-\frac{2}{m^2}N^\pr)g_{\mu \mup}+ \frac{4}{m^2}r^2N^{\pr \pr}
n_\mu n_\nup, \\
\end{eqnarray*}
where $r^2=(x-y)^2$ and 
\[
N(r^2)=\int \frac{d^{d+1}k}{(2\pi)^{\frac{d+1}{2}}} \frac{e^{ik.(x-y)}}{k^2+m^2}.
\] 
As $u \rightarrow 0$, this approaches 
\beq
G^{flat}_{\mu \mup}=-\frac{2}{m^2}{\cal N}^\pr(r^2) g_{\mu \mup}+ 
\frac{4}{m^2}r^2{\cal N}^{\pr \pr}(r^2)n_\mu n_\nup.
\eeq
${\cal N}$
 is the leading term in $N(r^2)$ as $r^2 \rightarrow 0$\footnote{${\cal N} \sim \frac{1}{r^{d-1}}$} .
Also for small $u$, $u \approx \frac{\mu^2}{2}=\frac{r^2}{2}$ 
(from Eq(\ref{dictionary})). 

The normalization of $F(u)$ in Eq(\ref{eq539}) was chosen such that
it matches with $N(r^2)$ as $u \rightarrow 0$, i.e.
\[
{\cal F}(u)= {\cal N}(r^2)={\cal N}(2u).
\]
In terms of $u$, as $u \rightarrow 0$, 
\beq
G^{flat}_{\mu \mup}=-\frac{1}{m^2}\frac{d}{du}{\cal N}(2u) g_{\mu \mup}+ 
\frac{2}{m^2}u \frac{d^2}{du^2}{\cal N}(2u)n_\mu n_\nup,
\eeq 
which is exactly the same as the short distance limit in Eq(\ref{sdlvector}).
\subsection{Massive symmetric tensor}
As $ u \rightarrow 0 $, $G_{\mu \nu; \mup \nup}(z,w)$ for the massive
symmetric tensor should approach
the short-distance limit of the massive symmetric tensor propagator in flat space. To 
check this, it is convenient re-write our expression for 
$G_{\mu \nu; \mup \nup}(z,w)$ in terms of tensors ${\cal O}^{i}$'s defined
in \cite{dfmmr}:
\begin{eqnarray} \label{O's}
{\cal O}_{\mu \nu  \mu' \nu'}^{(1)} &=& g_{\mu \nu}g_{\mu' \nu'}, \nonumber \\
{\cal O}_{\mu \nu  \mu' \nu'}^{(2)} &=& n_\mu n_\nu n_{\mu'} n_{\nu'}, \nonumber \\
{\cal O}_{\mu \nu  \mu' \nu'}^{(3)} &=& g_{\mu \mu'} g_{\nu \nu'}+
                                  g_{\mu \nu'} g_{\nu \mu'},\\
{\cal O}_{\mu \nu  \mu' \nu'}^{(4)} &=& g_{\mu \nu} n_{\mu'} n_{\nu'}+ 
g_{\mu' \nu'} n_{\mu} n_{\nu}, \nonumber \\
{\cal O}_{\mu \nu  \mu' \nu'}^{(5)} &=& g_{\mu \mu'} n_{\nu} n_{\nu'}+g_{\mu \nu'} n_{\nu} n_{\mu'} + g_{\nu \nu'} n_{\mu} n_{\mu'} +g_{\nu \mu'} n_{\mu} n_{\nu'} \,.
\nonumber
\end{eqnarray}
 We can write the tensors $T_i$ 
defined earlier in terms of ${\cal O}_i$ by using Eq(\ref{dictionary}):
\begin{eqnarray*}
T^{(1)}&=&{\cal O}^{(1)}, \\
T^{(2)}&=&[u(u+2)]^2{\cal O}^{(2)} \, ,\\
T^{(3)}&=&2u^2{\cal O}^{(2)}+{\cal O}^{(3)}-u{\cal O}^{(5)} \, ,\\
T^{(4)}&=&u(u+2){\cal O}^{(4)}\, ,\\
T^{(5)}&=&4u^2(u+2){\cal O}^{(2)}-u(u+2){\cal O}^{(5)}\, .\\
\end{eqnarray*}
$G_{\mu \nu; \mup \nup}(z,w)$ can be written in terms of ${\cal O}^{(i)}$:
\begin{eqnarray*}
G_{\mu \nu; \mup \nup}(z,w) &= & \Bigl(H+4(1+u)A \Bigr){\cal O}^{(1)}+
\Bigl(G+4B\Bigr){\cal O}^{(3)} \\
&+&\Bigl(4u^2(u+2)^2C^\pr +2u^2G+8u^2B+8u^2(u+2)C+8u^2(u+2)B^\pr \Bigr) {\cal O}^{(2)}
 \\
& + & \Bigl(2A^\pr+2(1+u)C+4B\Bigr)u(2+u){\cal O}^{(4)} \\ &-&
\Bigl( (G+4B)u+(2C+2B^\pr)u(2+u)\Bigr) {\cal O}^{(5)} \, . 
\end{eqnarray*}
We now take the $u \rightarrow 0$ and $m \rightarrow \infty$ limit:
\begin{eqnarray*}
G_{\mu \nu; \mup \nup}(z,w) &= & \frac{1}{2dm^4}\Bigl[
4(d-1){\cal G}^{\pr \pr} {\cal O}_1+
4(d-1){\cal G}^{\pr \pr} {\cal O}_3 +16(d-1)u^2{\cal G}^{\pr \pr \pr \pr} {\cal O}_2 \\
 &+& 8(d-1)u{\cal G}^{\pr \pr \pr}{\cal O}_4- 8(d-1)u{\cal G}^{\pr \pr \pr} {\cal O}_5  
\Bigr] \, .
\label{sdlimit}
\end{eqnarray*}
Here, ${\cal G}$ is the leading term in $G(u)$ as
$u\rightarrow 0$\footnote{${\cal G} \sim \frac{1}{u^{\frac{d-1}{2}}}$.}.

In flat Euclidean space, the propagator for the massive symmetric tensor is calculated
in the appendix. The short distance limit of the flat space propagator is
given by the expression: 
\begin{eqnarray*}
G^{flat}_{\mu \nu; \mup \nup}(x,y) & = &
\frac{8(d-1)}{m^4d}{\cal N}^{\pr \pr}  {\cal O}^{(1)}
+\frac{32(d-1)}{m^4d}r^4 {\cal N}^{\pr \pr \pr \pr} {\cal O}^{(2)} 
 +   \frac{8(d-1)}{m^4d}{\cal N}^{\pr \pr} 
{\cal O}^{(3)}\\&+& \frac{16(d-1)}{m^4d}r^2
{\cal N}^{\pr \pr \pr} {\cal O}^{(4)}   -   \frac{16(d-1)}{m^4d}r^2
{\cal N}^{\pr \pr \pr}  {\cal O}^{(5)} \, .
\label{flatsdl}
\end{eqnarray*}
where ${\cal N}$ is, as before, the leading term in $N(r^2)$. 
$G(u)$ was normalized in Eq(\ref{normg}) such that
\[
{\cal G}(u)={\cal N}(r^2)={\cal N}(2u) \, ,
\]
as $u=\frac{r^2}{2}\rightarrow 0$. Then, in terms of $u$, we can
write Eq(\ref{flatsdl}) as
\begin{eqnarray*}
G^{flat}_{\mu \nu; \mup \nup}(z,w) &= & \frac{1}{2dm^4}\Bigl[
4(d-1)\frac{d^2}{du^2}{\cal N}(2u) {\cal O}^{(1)}+
4(d-1)\frac{d^2}{du^2}{\cal N}(2u) {\cal O}^{(3)}\\ &+&16(d-1)u^2
\frac{d^4}{du^4}{\cal N}(2u) {\cal O}^{(2)} 
 + 8(d-1)u \frac{d^3}{du^3}{\cal N}(2u){\cal O}^{(4)} \\ &-& 8(d-1)u
\frac{d^3}{du^3}{\cal N}(2u) {\cal O}^{(5)} \, 
\Bigr].
\end{eqnarray*}
This exactly matches the short distance limit of the propagator
in $AdS_{d+1}$ in Eq(\ref{sdlimit}). 
\section{Summary of results}
We have calculated the propagators for the massive symmetric tensor and
$p$-form fields in $AdS_{d+1}$ using the method developed in \cite{dfmmr}.
In this section, we summarize our results. 
\newline
\rule{6.35in}{0.005in}
\subsection*{$p$-forms}
We defined two independent bi-tensors:
\begin{eqnarray*}
T_{\mo \mt \dots \msp}{}^{ \mop \mtp \dots \mpp} & =& 
 \de_{[\mo} \de^{\mop} u \de_\mt \de^\mtp u \dots \de_{\msp]} \de^{\mtp}u 
\, , \\
S_{\mo \mt \dots \msp}{}^{ \mop \mtp \dots \mpp} & =& 
 \de_{[\mo} u\de^{[\mop} u \de_\mt \de^\mtp u \dots \de_{\msp]} \de^{\mtp]}u \, .
\end{eqnarray*}
Then, 
\[
G_{\mo \mt \dots \msp}{}^{ \mop \mtp \dots \mpp}=(F(u)+pL(u)) 
T_{\mo \mt \dots \msp}{}^{ \mop \mtp \dots \mpp}+L^\pr(u) 
S_{\mo \mt \dots \msp}{}^{ \mop \mtp \dots \mpp} \, .
\]
\begin{eqnarray*}
F(u) & = & \tilde{C}_{\Delta} (2u^{-1})^\Delta F(\Delta, \Delta-\frac{d}{2}+\frac{1}{2}; 2\Delta -d+1;
-2u^{-1}) \, ,\\
 \tilde{C}_{\Delta} & = & \frac{\Gamma(\Delta)\Gamma(\Delta-\frac{d}{2}+\frac{1}{2})}
{(4 \pi )^{(d+1)/2} \Gamma(2 \Delta -d +1)}\, ,
\end{eqnarray*}
where
\[
\Delta=\frac{d}{2}+\frac{1}{2}\sqrt{d^2+4m^2-4p(d-p)} \, ,
\]
\[
L(u)=-\frac{1}{m^2}\Bigl((d-p)F(u)+(1+u)F^\pr(u) \Bigr) \, .
\]
\newline
\rule{6.35in}{0.005in}
\subsection*{Massive symmetric tensor}
In terms of  $T^{(i)}$ defined in Eq(\ref{T's}),
\begin{eqnarray*}
G_{\mu \nu; \mup \nup}(z,w)&=&\Bigl(H(u)+4(1+u)A(u) \Bigr) T^{(1)}_{\mu \nu; \mup \nup}
+4C^\pr(u)T^{(2)}_{\mu \nu; \mup \nup}+\Bigl( G(u)+4B(u) \Bigr) 
T^{(3)}_{\mu \nu; \mup \nup} \\
& + & \Bigl( 2A^\pr(u)+2(1+u)C(u)+4B(u)\Bigr) T^{(4)}_{\mu \nu; \mup \nup}
+\Bigl( 2C(u)+2B^\pr(u)\Bigr) T^{(5)}_{\mu \nu; \mup \nup} \,.
\end{eqnarray*}
This can also be written in terms of ${\cal O}^{(i)}$ (Eq(\ref{O's}), 
\begin{eqnarray*}
G_{\mu \nu; \mup \nup}(z,w) &= & \Bigl(H+4(1+u)A \Bigr){\cal O}^{(1)}+
\Bigl(G+4B\Bigr){\cal O}^{(3)} \\
&+&\Bigl(4u^2(u+2)^2C^\pr +2u^2G+8u^2B+8u^2(u+2)C+8u^2(u+2)B^\pr \Bigr) {\cal O}^{(2)}
 \\
& + & \Bigl(2A^\pr+2(1+u)C+4B\Bigr)u(2+u){\cal O}^{(4)} \\ &-&
\Bigl( (G+4B)u+(2C+2B^\pr)u(2+u)\Bigr) {\cal O}^{(5)}  \,.
\end{eqnarray*}
We found, 
\[
G(u)  =  \tilde{C}_{\Delta} (2u^{-1})^\Delta F(\Delta, \Delta-\frac{d}{2}+\frac{1}{2}; 2\Delta -d+1;
-2u^{-1}) \, ,
\]
where
\[
\Delta=\frac{d}{2}+\frac{1}{2}\sqrt{d^2+4m^2} \, .
\]
\[
C(u)=\frac{1}{2d(d-1+m_1^2)m_1^2}\Bigl((d-1)G'''+d(d-1)(1+u)G''+(d^3-d+m_1^2d)G'\Bigr) \, ,
\]
\[
B(u)=\frac{1}{2dm_1^2(d-1+m_1^2)}\Bigl((d-1)G''-m_1^2d (1+u)G'-m_1^2d(d-1)G\Bigr) \, ,
\]
\begin{eqnarray*}
A(u)&=& \frac{1}{2m_1^2d(d-1+m_1^2)}\Bigl((d-1)(1+u)G''+(d^2-1+2m_1^2)G'\nonumber \\
&& +m_1^2d(1+u)G + m_1^2 d(d-2) \int G \Bigr) \, ,
\end{eqnarray*}
\[
H(u)=-\frac{2}{d(d-1+m_1^2)}\Bigl((2d^2+m_1^2d-4d) \int \int G + d(1+u)\int G + (d+m_1^2)G\Bigr) \, .
\]
\newline
\rule{6.35in}{0.02in}

\begin{center}{\bf Acknowledgments}\end{center}
Its a pleasure to thank Dan Freedman and Leonardo Rastelli
for several helpful discussions. We also thank Dan Freedman
for suggesting the project and for comments on the manuscript. 
This research is supported in part by the U.S. Department of Energy 
under cooperative agreement \#DE-FC02-94ER40818.

\newpage
\appendix
\section{Appendix}
\section*{Massive symmetric tensor propagator in $R^{d+1}$}
\setcounter{equation}{0}
In this section, we will calculate the graviton propagator in flat
space. The equation of motion is
\beq
-\de^{\sigma}\de_\sigma S_{\mu \nu}-\dmu \dnu S_{\sigma}{}^{\sigma}
+\dmu \de^{\sigma}S_{\sigma \nu}+\dnu \de^{\sigma}S_{\sigma \mu}+m_1^2 S_{\mu \nu}
+m_2^2 g_{\mu \nu}S_{\sigma}{}^{\sigma} = \tilde{T}_{\mu \nu} \,.
\label{flatlinear}
\eeq
where $\tilde{T}_{\mu \nu}= {T}_{\mu \nu}-\frac{1}{d-1}g_{\mu \nu}T_\sigma{}^\sigma$. 
The Green's function is defined by the equation
\[
S_{\mu \nu}=\int d^{d+1}y G_{\mu \nu;\mup \nup}(x,y)T^{\mup \nup}(y) \,.
\]
We will work in momentum space. Let
\[
S_{\mu \nu}=\int \frac{d^{d+1}k}{(2 \pi)^{\frac{d+1}{2}}} s_{\mu \nu}(k)e^{ikx},
\]
and
\[
T_{\mu \nu}=\int \frac{d^{d+1}k}{(2 \pi)^{\frac{d+1}{2}}} t_{\mu \nu}(k)e^{ikx}.
\]
Then
\[
s_{\mu \nu}=G_{\mu \nu;\mup \nup}(k)t^{\mup \nup}.
\]
In momentum space Eq(\ref{flatlinear}) becomes: 
\begin{eqnarray}
& & k^2 G_{\mu \nu;\mup \nup}+k_\mu k_\nu G_{\sigma}{}^\sigma{}_{;\mup \nup}
-k_\mu k^\sigma G_{\sigma \nu;\mup \nup}-k_\nu k^\sigma G_{\sigma \mu;\mup \nup}
+m_1^2 k_\mu k^\sigma G_{\sigma \nu;\mup \nup}+
m_2^2g_{\mu \nu}G_{\sigma}{}^\sigma{}_{;\mup \nup}  \nn \\
&&=g_{\mu \mup}g_{\nu \nup}
+g_{\nu \mup}g_{\mu \nup}-\frac{2}{d-1}g_{\mu \nu}g_{\mup \nup}.
\label{flat}
\end{eqnarray}
We will use the following ansatz for $G_{\mu \nu;\mup \nup}$:
\begin{eqnarray*}
G_{\mu \nu;\mup \nup} &=&A(k^2)g_{\mu \nu}g_{\mup \nup}+B(k^2)(g_{\mu \mup}g_{\nu \nup}
+g_{\nu \mup}g_{\mu \nup})+C(k^2)(g_{\mu \nu} k_{\mup}k_{\nup}+
g_{\mup \nup} k_{\mu}k_{\nu}) \\
& + & F(k^2)(g_{\mu \nup}k_\mup k_\nu +g_{\nu \nup}k_\mup k_\mu + 
g_{\mu \mup}k_\nu k_\nup +g_{\nu \mup}k_\mu k_\nup )+G(k^2)k_\mu k_\mup k_\nu k_\nup \, .
\end{eqnarray*}
Using this ansatz in Eq(\ref{flat}), we get
\begin{eqnarray}
& &\Bigl( (k^2+m^2+m_2^2(d+1))A(k^2) + 2 m_2^2 B(k^2)+m_2^2k^2C(k^2) \Bigr)
g_{\mu \nu}g_{\mup \nup}\nn \\ & +& (k^2+m^2)B(k^2)(g_{\mu \mup}g_{\nu \nup}
+g_{\nu \mup}g_{\mu \nup}) \nn \\
& +&\Bigl( (k^2+m_1^2+(d+1)m_2^2) C(k^2) + 4m_2^2 F(k^2)+m_2^2 k^2 G(k^2) \Bigr)
g_{\mu \nu}k_{\mup}k_{\nup}  \nn \\
&+&  \Bigl( m_1^2 C(k^2)+(d-1)A(k^2)+2B(k^2)\Bigr)g_{\mup \nup}k_{\mu}k_{\nu} \nn \\
&+&\Bigl( m_1^2F(k^2)-B(k^2)\Bigr)(g_{\mu \nup}k_\mup k_\nu +g_{\nu \nup}k_\mup k_\mu + 
g_{\mu \mup}k_\nu k_\nup +g_{\nu \mup}k_\mu k_\nup ) \nn \\
& + & \Bigl( m_1^2G(k^2)+(d-1)C(k^2) \Bigr)k_\mu k_\mup k_\nu k_\nup = (g_{\mu \mup}g_{\nu \nup}
+g_{\nu \mup}g_{\mu \nup}) -\frac{2}{d-1}g_{\mu \nu}g_{\mup \nup} \nn \, .
\end{eqnarray} 
This gives a system of linear equations for the functions $A(k^2),B(k^2),C(k^2),F(k^2),G(k^2)$ which can be easily solved:
\[
A(k^2)=-\frac{2}{d(k^2+m^2)};B(k^2)=\frac{1}{k^2+m^2}; C(k^2)=-\frac{2}{m^2d(k^2+m^2)}
\]
\[
F(k^2)=\frac{1}{m^2(k^2+m^2)};G(k^2)=\frac{2(d-1)}{m^4d(k^2+m^2)}.
\]
In position space, defining $r^2=(x-y)_\mu(x-y)^\mu$, 
\begin{eqnarray*}
G_{\mu \nu;\mup \nup}(x,y)& = & \Bigl(-\frac{2}{d}g_{\mu \nu}g_{\mup \nup} + g_{\mu \mup}g_{\nu \nup}
+g_{\nu \mup}g_{\mu \nup}+\frac{2}{m^2d}(g_{\mu \nu} \de_{\mup}\de_{\nup}+
g_{\mup \nup} \de_{\mu}\de_{\nu})\nn \\ & + & \frac{1}{m^2}(g_{\mu \nup}\de_\mup \de_\nu +g_{\nu \nup}\de_\mup \de_\mu + 
g_{\mu \mup}\de_\nu \de_\nup +g_{\nu \mup}\de_\mu \de_\nup ) \\ &+&\frac{2(d-1)}{m^4 d}
\de_\mu \de_\mup \de_\nu \de_\nup \Bigr)N(r^2) \\
& = &  \Bigl(-\frac{2}{d}+\frac{8(d-1)}{m^4d}N^{\pr \pr}+\frac{8}{m^2d} N^\pr \Bigr) {\cal O}^{(1)}
+\frac{32(d-1)}{m^4d}r^4 N^{\pr \pr \pr \pr} {\cal O}^{(2)} \\
& + &  \Bigl( 1-\frac{4}{m^2}N^{\pr}+ \frac{8(d-1)}{m^4d}N^{\pr \pr} \Bigr)
{\cal O}^{(3)}+\Bigl( \frac{8}{m^2d} r^2 N^{\pr \pr}+\frac{16(d-1)}{m^4d}r^2
N^{\pr \pr \pr} \Bigr){\cal O}^{(4)} \\ & + &\Bigl( \frac{4}{m^2}r^2 
N^{\pr \pr}- \frac{16(d-1)}{m^4d}r^2
N^{\pr \pr \pr} \Bigr) {\cal O}^{(5)}.
\end{eqnarray*}
Here, ${\cal O}^{(i)}$'s are defined in Eqs(\ref{O's}). In the flat space case, 
$n_\mu=\frac{(x-y)_\mu}{r}$ and $n_\mup=-\frac{(x-y)_\mup}{r}$.

\end{document}